\documentstyle{elsart}

\newcommand{\be}{\begin{equation}}
\newcommand{\ee}{\end{equation}}
\newcommand{\bee}{\begin{eqnarray}}
\newcommand{\eee}{\end{eqnarray}}
\begin{document}
\begin{frontmatter}
\title
      { Near threshold $\eta$ production in proton-proton collisions}
      \author[Rudjer]{M.  Batini\'{c}, A.  \v{S}varc\thanksref{US-CRO}}
\author[Argonne]{and T.-S. H. Lee\thanksref{US-CRO}}
\address[Rudjer]{\em Rudjer Bo\v{s}kovi\'{c} Institute,  Zagreb,
 Croatia }
\address[Argonne]{\em Physics Division, 
Argonne National Laboratory, IL 60439, USA}
\thanks[US-CRO]{Partially supported by US-CRO contract JF 221}
 \begin{abstract}
      The  total  cross section for the near threshold $\eta$ production
      in  proton-proton  collisions  has  been  investigated  with   the
      assumption that the production mechanism is due to the emission of
      a meson $x(\pi, \eta, \sigma)$ from one of the  nucleons  followed
      by  a  $x N \rightarrow \eta N$ transition on the second one.  The
      $xN \rightarrow \eta N$ amplitudes are generated from the  unitary
      multi-channel   multi-resonance  model  which  has  recently  been
      constructed in  analyzing  the  $\pi  N$  scattering  and  $\pi  N
      \rightarrow   \eta  N$  reaction.   The  initial  and  final  $pp$
      distortions are calculated from a coupled-channel $\pi  NN$  model
      which  describes the $NN$ scattering data up to about 2 GeV.  With
      the $x-NN$ vertex functions taken from  the  Bonn  potential,  the
      predicted  total  cross sections of threshold $pp \rightarrow \eta
      pp$ reaction are in good agreement with the data.  In contrast  to
      previous works, we find that the $\eta$-exchange plays an important
      role.  The effect of the two-pion exchange, simulated by $\sigma$-
      exchange, is found to be significant but not as dominant as the
      vector meson-exchange introduced in previous works.

\end{abstract}
\end{frontmatter}

      The  $\eta$ production in proton-proton collisions has attracted a
      lot of interest in the past decade.  A  theoretical  understanding
      of  this  two-nucleon  process  near  the  production threshold is
      needed for exploring the $N^*(S_{11}(1535))$ dynamics  in  nuclear
      medium, possible existence of $\eta$-nucleus bound states, and the
      possibility  of  using  $\eta$  to  reveal   the   properties   of
      high-density  nuclear  matter  created  in  relativistic heavy-ion
      collisions.      All     of     the      existing      theoretical
      works\cite{ger89,ger90,lag91,vet91,wil93}  on  this  reaction were
      based  on  the  assumption  that  the  basic  mechanism   of   $NN
      \rightarrow   \eta  NN$  reaction  is  the  emission  of  a  meson
      $x(\pi,\eta,\rho,\cdot\cdot)$ from one of the nucleons followed by
      a  $xN  \rightarrow  \eta  N$ transition, as illustrated in Fig.1.
      Clearly, an  accurate  assessment  of  this  meson-exchange  model
      depends  heavily  on  the accuracy of the employed $xN \rightarrow
      \eta N$ amplitudes, treatment of the propagation of the  exchanged
      mesons,  and  the  meson$-NN$  vertex functions.  Furthermore, the
      initial and final $NN$ interactions must be accounted for by using
      realistic  NN models.  In this work, we investigate      all these
      problems  by making use of the recent advance in constructing  the
      $xN  \rightarrow  \pi  N$  amplitudes\cite{bat95a,bat95b}  and the
      knowledge we have learned from previous studies of  meson-exchange
      models\cite{bonn,lee} of nucleon-nucleon interactions.

      A        common        feature        of        the       previous
      investigations\cite{ger89,ger90,lag91,vet91,wil93}     of      the
      threshold  $pp\rightarrow  \eta  pp$ reaction is that the $\eta N$
      channel is only due to the decay of  the  $N^*(S_{11})$  resonance
      and  the  agreement  with  the  data  is  obtained  only  when the
      subthreshold  exchange  of  vector  mesons  is  introduced.    The
      predicted  results thus depend heavily on how the assumptions were
      made to determine the vector meson coupling  constants  associated
      with    the    decay   of   the   $N^*(S_{11})$   resonance.    In
      Refs.\cite{ger89,ger90,lag91,wil93}, the  vector  meson  dominance
      model(VDM) is used to determine the $N^*(S_{11})- \rho N$ coupling
      from   the   branching   ratio    $\Gamma(N^*\rightarrow    \gamma
      N)/\Gamma(N^*\rightarrow \pi N) =0.013\pm 0.004$ \cite{ger89}.  In
      Ref.\cite{vet91},  this  $\rho$  coupling  constants  is  directly
      determined  from  the  $N^*(S_{11})\rightarrow \rho N$ decay width
      listed  by  the  Particle  Data  Group\cite{pdg}.   The   $\omega$
      exchange  is  also introduced in Ref.\cite{vet91} by assuming that
      its coupling with the $N^*$ relative to that to  the  nucleon  has
      the      same     value     as     for     the     pion;      i.e.
      $g_{NN^*\omega}/g_{NN\omega} =  g_{NN^*\pi}/g_{NN\omega}$.   These
      differences  in  choosing  the vector meson coupling constants had
      led   to   conflicting   conclusions   between    these    earlier
      investigations.   The  $\rho$-exchange was found to be dominant in
      Refs.\cite{ger89,ger90,lag91},  while  the   importance   of   the
      $\omega$-exchange was stressed in Ref.\cite{vet91}.

      In  this  work,  we  will  not  follow these earlier approaches to
      consider the possible contributions from vector mesons  exchanges.
      Instead,  our main objective is to examine the extent to which the
      threshold $pp \rightarrow \eta pp$ data can be understood from the
      $\pi$  and  $\eta$  exchange  which  can  be  calculated  with  no
      adjustable parameters.  This will be  accomplished  by  using  the
      $\pi  N  \rightarrow \eta  N$  and  $\eta  N$  elastic  amplitudes
      generated from the  recently  constructed  unitary  multi-channel,
      multi-resonance model of Ref.\cite{bat95a,bat95b} and the $\pi-NN$
      and   $\eta-NN$   vertex   functions   taken   from    the    Bonn
      potential\cite{bonn}.   We  then  estimate  the  importance of the
      two-$\pi$ exchange by assuming that the  fictitious  third-channel
      in  the employed unitary multi-channel, multi-resonance model is a
      $\sigma N$ channel with $\sigma$ identified with the  one  in  the
      same Bonn potential.  Another important feature of our calculation
      is the $\pi NN$ model developed in Ref.  \cite{lee} which accounts
      for  the  initial  and  final  $pp$  distortions.   This model was
      constructed to give a good description of $NN$  scattering  up  to
      about  2  GeV.  The use of such a realistic model is essential for
      calculating the initial $pp$ distortion in the  considered  rather
      high  energies  $E_L  \sim 1.3$ GeV.  The final $pp$ distortion is
      also calculated from the same model which is as  accurate  as  the
      Paris potential at low energies, as discussed in
Ref.\cite{lee}.
       \begin{center}
       \begin{tabular}{ccc}
       \hline
       &Table I& \\
        $x$ &$\frac{g_{xNN}^2}{4\pi}$&  $\Lambda_x$ ( MeV/c)      \\
       \hline
        $ \pi   $ &  14.9 & 1300 \\
        $ \eta  $ &   2.0 & 1500 \\
        $ \sigma $ &   8.7 & 2000 \\
        \hline
        \end{tabular}
        \end{center}
      In terms of the kinematical variables in Fig.1, the expression for
      the  spin-averaged  total  cross section of $pp\rightarrow pp\eta$
      reaction can be written as the following form :
        \be
           \sigma_{tot} = \int dq d\Omega_{\eta} d\Omega_{1}
           \frac{d^{5}\sigma}{dq d \Omega_{\eta} d\Omega_{1}} ,
        \ee
      with
        \bee
           \frac{d^{5}\sigma}{dq d \Omega_{\eta} d\Omega_{1}}&=&
           \frac{ 2m^4
           q^{2} p_{1}^{2}}
           {(2\pi)^{5} E_{\eta}(p_{1}E_{2}-E_{1}\vec{p}_{1}
           \hat{p}_{2})} \frac{1}{\sqrt{ (p_b\cdot p_t)^{2}-m^{4} }}
           [\frac{1}{4}\sum_{spins} | M_{fi}|^2] .
        \eee
      with $p_b$ and $p_t$ being beam and target four momenta, and $p_1$
      and $p_2$ ($E_1$ and $E_2$) are the final nucleon  three  c.m.
      momenta  and  energy.  $E_{\eta}$ is the $\eta$-c.m.  energy.  The
      transition matrix element is defined by
\be
M_{fi} = <\chi_f^{(-)}|  M^{0} | \chi^{(+)} >,
\ee
      where $\chi_f^{(-)}$ and $\chi_i^{(+)}$ are respectively the  final
      and  initial $pp$ scattering wavefunctions generated from the $\pi
      NN$ model of Ref.\cite{lee}.  The production operator $M^0$ is
\be
M^0 = \sum_{x=\pi,\eta,\sigma} M^{x,0},
\ee
      where the contribution from each exchanged meson $x$ is defined by
      the following plane-wave invariant matrix element
        \bee
           M^{x,0} & = &
           \left\{ \left[ \overline{u}(p_1,s_1)(A_x + \not \! q B_x) u(p_b,s_b)
           \frac{i}{p_x^2 - m_x^2}\overline{u}(p_2,s_2)V_{xNN}(p^2_x)u(p_t,s_t)
           \right] \right. \nonumber \\
           & - & \left. \left[ 1 \leftrightarrow 2 \right] \right\} \\
           &-&  \left\{ b \leftrightarrow t \right\}. \nonumber
        \eee
      Here,  $A_x$  and  $B_x$ are the $xN \rightarrow \eta N$ invariant
      functions    generated    from    the    unitary    multi-channel,
      multi-resonance   model   of   Ref.\cite{bat95a,bat95b}.    It  is
      worthwhile  to  note       that  the  range  of  the  constructed
      production  operator depends on how the momentum-transfer variable
      $p_x$ in Eq.(5) is evaluated.  In this work, we follow  the  usual
      meson-exchange  model  of  nuclear force\cite{bonn,lee} to set all
      nucleons  on  their  mass-shell  to  choose  $p^2_x=(p_t-p_2)^2  =
      (E_N(\vec{p}_t)-E_N(\vec{p}_2))^2  -(\vec{p}_t-\vec{p}_2)^2$.  The
      resulting operator is always finite  range  in  coordinate  space.
      This     is    very    different    from    the    treatment    of
      Ref.\cite{ger89,ger90,wil93}.    These   authors    obtained    an
      oscillating   one-pion-exchange  which  is  of  infinite  range  in
      coordinate space.  

      The  $x-NN$  vertex  function  in  Eq.(5)  is  parameterized  as a
      monopole form
      \be
          V_{xNN} = g_{xNN}\frac{\Lambda_x^2 - m_x^2}{\Lambda_x^2 - p_x^2},
      \ee
      The parameters $\Lambda_x$ and $g_{xNN}$ used  in  this  work  are
      taken  from  the  Bonn  potential\cite{bonn} and is given in Table
      1(from Table A.3 of Ref.\cite{bonn}).

      We first examine the distortion effects.  In Fig.2, we display the
      distortion factors calculated from the following expressions:
\begin{eqnarray}
f_{ISI} &=& \frac{\int N|<\chi_f^{(-)} | M^{0}  |\chi_i^{(+)} >|^2 d X}  
                 {\int N|<\chi_f^{(-)} | M^{0}  |\phi_i> |^2 d X} \\
f_{FSI} &=& \frac{\int N|<\chi_f^{(-)} | M^{0}  |\phi_i>|^2 d X}
                 {\int N|<\phi_f | M^{0} | \phi_i >^2 d X} \\
f_{tot}&=& f_{FSI}\cdot f_{ISI}
\end{eqnarray}
      where  $d  X = dq d \Omega_{\eta} d \Omega_1$ and $N$ contains all
      of the kinematic factors in Eq.(2).   $\phi_i$  and  $\phi_f$  are
      plane  waves.  Clearly these factors measure the importance of the
      $pp$ distortion in the $\eta$ production.  The initial  distortion
      factor   $f_{ISI}$(dotted   curve)  is  mainly  due  to  the  $pp$
      interaction in the $^3P_0$ channel.  It reduces the  magnitude  of
      the cross sections by about 40 percent at all energies considered.
      On  the  other  hand,  the  final   distortion   $f_{FSI}$(dashed)
      drastically  increases  the cross section as the energy approaches
      the threshold.  This is due to the formation of a quasi-bound $pp$
      state  in  the  $^1S_0$  channel  at very low energies.  The total
      distortion $f_{tot}$ increases the plane-wave cross sections by  a
      factor  of  about  1.5  at  energies  close  to the threshold, but
      reduces the  cross  section  at  higher  energies.   Clearly,  the
      initial  and  final  state  interactions  play nontrivial roles in
      determining the calculated cross sections.  This is  however  well
      known from many nuclear reaction calculations using distorted-wave
      impulse approximation\cite{feshbach}.  The results shown in  Fig.2
      already  distinguish our work from all previous works in which the
      initial  $pp$  distortion  at  such  high   energies   is   either
      neglected\cite{lag91}   or   evaluated\cite{ger89,ger90,wil93}  by
      using a $NN$ potential which is only valid at  low  energies.   In
      Ref.\cite{vet91},  the initial and final $pp$ distortions are both
      neglected.

      From the definitions Eqs.(7)-(9), we obviously can write Eq.(1) as
\be
\sigma_{tot} = f_{tot} \sigma^0_{tot} 
\ee
      where $\sigma^0_{tot}$ can be calculated from  Eqs.(1)-(5)  except
      that  the distorted waves $\chi^{(\pm)}$ in Eq.(3) are replaced by
      the plane  waves  $\phi$.   We  have  found  that  the  calculated
      distortion factor $f_{tot}$ is rather insensitive to the detail of
      the production operator $M^0$ as far as it  is  of  the  range  of
      nuclear  force.   In  actual  calculations, we calculate $f_{tot}$
      from the simple $\sigma$-exchange mechanism.  Eq.(10) is then used
      in  the  calculation with the full complexities of Eq.(5) included
      in the plane-wave calculation of $\sigma^0_{tot}$.

      Before we discuss our results, it is necessary to point  out  that
      the  relative  phase  between the $\pi N$, $\eta N$ and $\sigma N$
      channels can  not  be  fixed  within  the  unitary  multi-channel,
      multi-resonance  model\cite{bat95a,bat95b}.   To  be  specific, we
      therefore choose $(\pi N \rightarrow \eta N)/(\eta  N  \rightarrow
      \eta  N)  \propto  -1$  according  to  the  SU(3)  quark  model of
      $N^*(S_{11}) \rightarrow \pi N, \eta N$  developed  by  Arima  et.
      al\cite{arima}.   However,  the phase of the $\sigma N \rightarrow
      \eta N$ relative to other transition amplitudes can not  be  fixed
      within  the phenomenological approach we are taking.  We therefore
      present results with both  $+$  and  $-$  relative  signs  between
      $\sigma  N  \rightarrow  \eta  N$  and  $\pi N \rightarrow \eta N$
      amplitudes.

      We first compare the contributions  from  the  $\pi$-exchange  and
      $\eta$-exchange.   Their  contributions are the full thin line and
      the full thicker lines in Fig.3.   In  contrast  to  all  previous
      works\cite{ger89,ger90,lag91,vet91,wil93},   they  are  comparable.
      The origin  of  this  is  that  the  $\eta  N$  scattering  length
      determined   in   the   unitary,   multi-channel,  multi-resonance
      model(see Table I of Ref.\cite{bat95b}) is about $0.88$  which  is
      considerably  larger  than  the  value $\le 0.55$ used in previous
      calculations\cite{ger89,ger90,lag91,vet91,wil93}.  This  important
      difference  was mainly due to the inclusion of multi-resonance and
      non-resonant background interaction in the fit to the data.   This
      was fully discussed in Ref.\cite{bat95b}.

      The  solid  curve  in  Fig.3 is obtained from our calculation with
      only  $\pi$-  and  $\eta$-exchange.   There   is   no   adjustable
      parameters    in    this    calculation.     By    including   the
      $\sigma$-exchange, we obtain the dashed and  dotted  curves.   The
      difference  between them is due to uncertainty of the phase of the
      $\sigma N\rightarrow \eta N$ transition relative  to  the  $\pi  N
      \rightarrow   \eta  N$  transition.   Clearly,  the  data  can  be
      described  within  the  theoretical  uncertainty  of  the  present
      approach.
 
      In  conclusion,  we  have  shown  that  the  near  threshold  $ pp
      \rightarrow \eta pp$ reaction can be described with no  adjustable
      parameters  by  using  a  meson-exchange  model  mediated  by  the
      exchange  of  $\pi,  \eta$  mesons.   The  use  of   the   unitary
      multi-channel,  multi-resonance model to generate the $\pi N, \eta
      N \rightarrow \eta N$ amplitudes is essential.   The  initial  and
      final  $pp$  distortions calculated from realistic $NN$ models are
      also shown to be very important in achieving the agreement with the
      data.   The  two-$\pi$ exchange, simulated by a $\sigma$-exchange,
      is found to be of some importance.  
      To  quantify  this
      mechanism,  it  is  necessary  to  investigate  the  corresponding
      mechanism  in  other  $\eta$  production  reactions  such  as  the
      photoproduction  of  $\eta$  meson on the nucleon.  Our effort in
      this direction will be reported elsewhere.

\newpage
\section*{Figure captions}
 \begin{description}
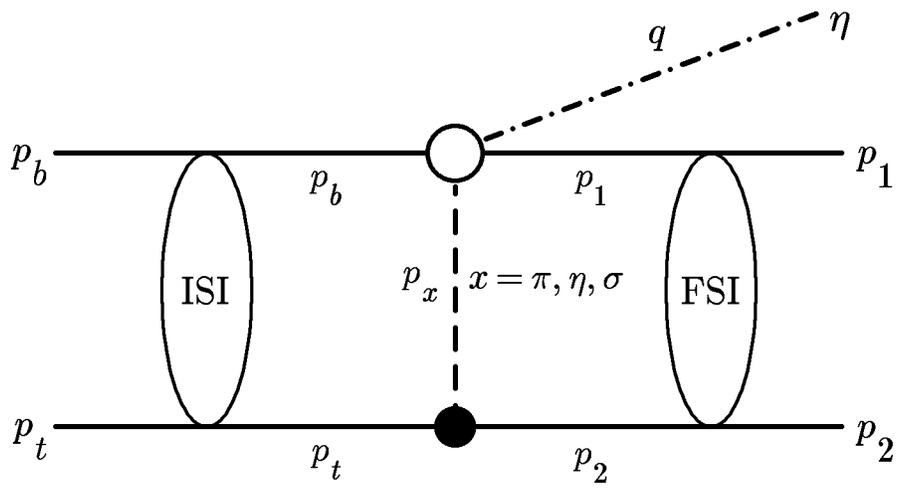
      \item[Fig.  1.] \hspace*{1.cm} \\
      The Feynman diagram describing our process. The description of the
      variables is given in the text.
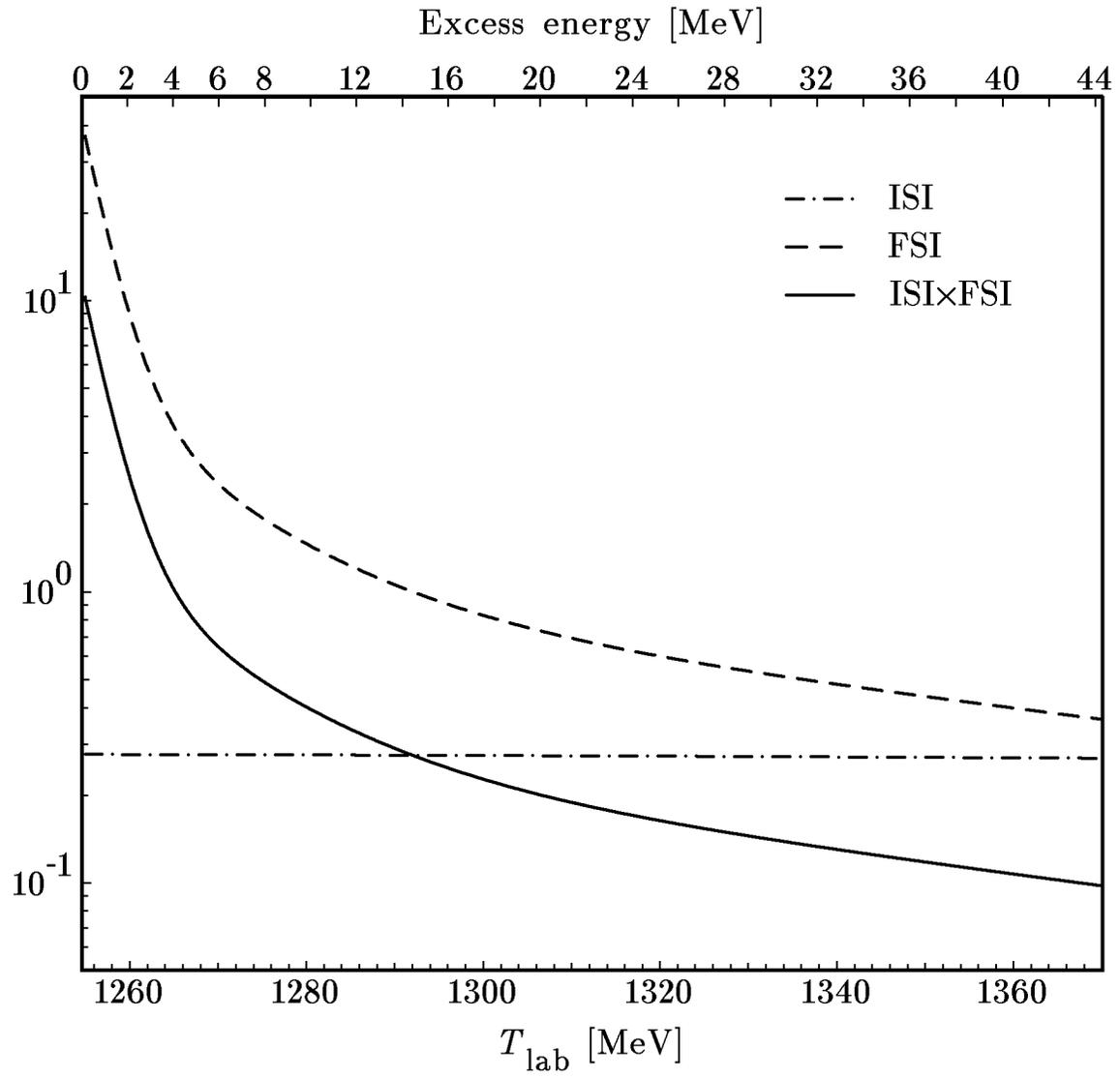
      \item[Fig. 2.] \hspace*{1.cm} \\
      The  values for the ISI and FSI distortion factors in the relevant
      energy range.  The exact definitions are given in the text.
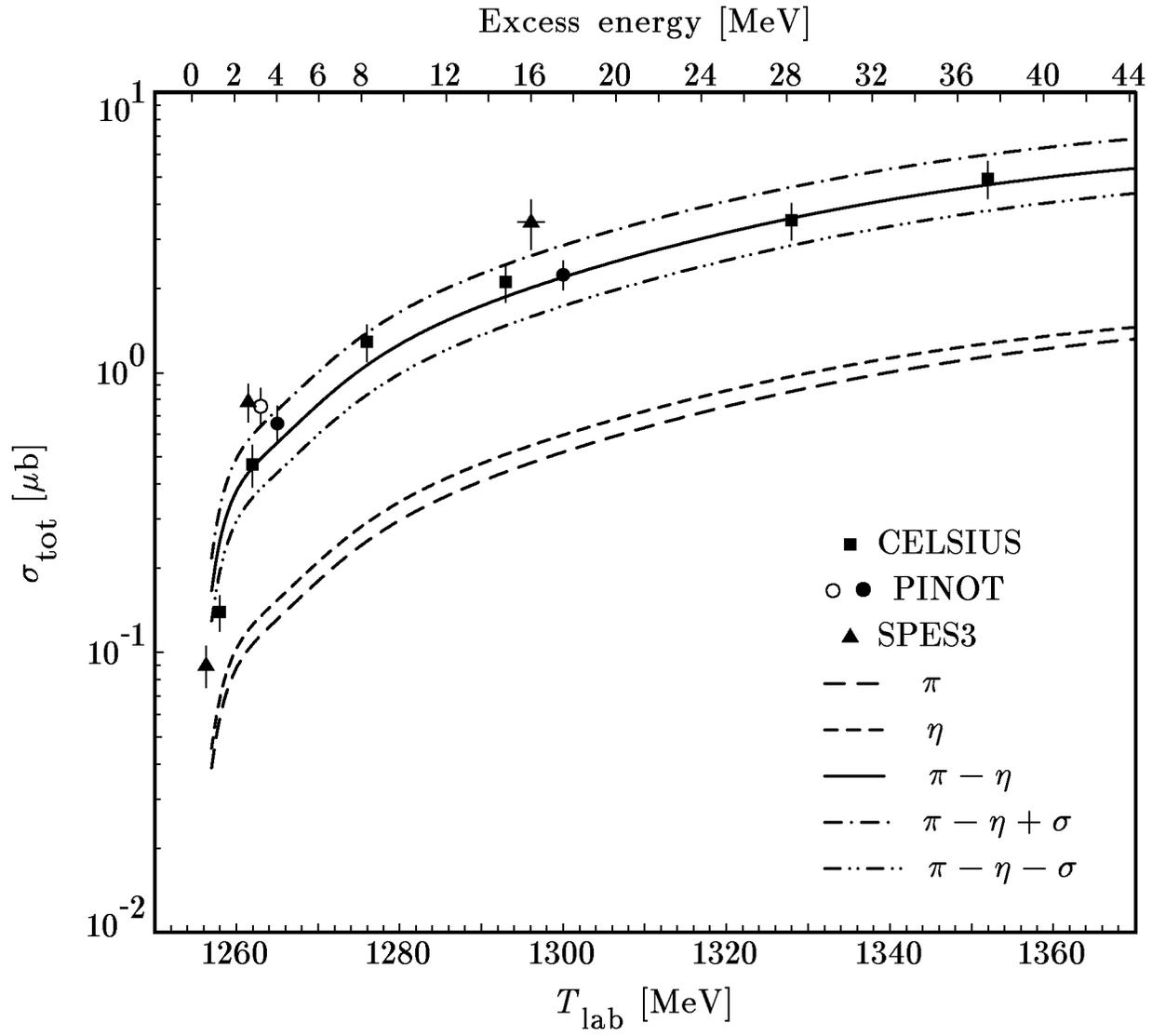
      \item[Fig. 3.] \hspace*{1.cm} \\
      The Comparison of near threshold experimental results for the
      $ pp \rightarrow  pp \eta$ process with the predictions of our
      model. The meaning of each line is given in the legend of the
      Figure.
\end{description}
\end{document}